\documentclass[showpacs,preprintnumbers,amsmath,amssymb,twocolumn]{revtex4}

\usepackage{graphicx}
\usepackage{dcolumn} 
\usepackage{bm}      

\begin{document}
\title{  Thermodynamic Properties of Small Localized  Black Holes }
\author{Hideaki Kudoh} \email{kudoh@yukawa.kyoto-u.ac.jp}
\affiliation{  Department of Physics, Kyoto University, Kyoto 606-8502, Japan  } 

\begin{abstract}
In a previous paper, we developed a numerical method to obtain a static black hole localized on a 3-brane in the Randall-Sundrum infinite braneworld, and presented examples of numerical solutions that describe small localized black holes. In this paper we quantitatively analyze the behavior of the numerically obtained black hole solutions, focusing on thermodynamic quantities. The thermodynamic relations show that the localized black hole deviates smoothly from a five-dimensional Schwarzschild black hole, which is a solution in the limit of a small horizon radius. We compare the thermodynamic behavior of these solutions with that of the exact solution on the 2-brane in the 4D braneworld. We find similarities between them. 
\end{abstract}

\preprint{KUNS-1846}
\preprint{hep-th/0306067}

\maketitle
\section{Introduction}
Black holes in higher dimensions have been studied in many publications.
A great deal of attention has recently been focused on black holes in
compactified spacetime in the context of the braneworld. This has resulted from the interesting proposal for the possibility of black hole production at colliders \cite{Giddings:2001bu,Dimopoulos:2001hw} in the scenario of large extra dimensions~\cite{Arkani-Hamed:1998rs}.
Such black holes in a compactified spacetime have been discussed in many works.
Some properties of 4D Schwarzschild black holes in a compactified spacetime are considered in Refs.~\cite{Myers:rx} and \cite{Frolov:2003kd} (see also references therein).
The black brane is another type of black hole in a compactified spacetime, and its instability, which is known as the Gregory-Laflamme instability~\cite{Gregory:1993vy,Gregory:1994bj,Gregory:2000gf}, has been studied extensively by many authors~\cite{Horowitz:2001cz,Wiseman:2002zc,Choptuik:2003qd,Gregory:2001,Gubser:2001ac,Gubser:2000mm,Kol:2003ja}.

Black holes in a warped spacetime are also interesting.
In the Randall-Sundrum (RS) braneworld model \cite{Randall:1999ee,Randall:1999vf}, the tension of the brane and the cosmological constant in the bulk are not negligible, and it is difficult to construct black hole solutions \cite{Charmousis:2003wm}.
Based on the Gregory-Laflamme instability, however, physically realistic black holes in the RS infinite braneworld are conjectured to be localized on the brane~\cite{Chamblin:2000by}. (See Ref.~\cite{Gibbons:2002pq} for a recent discussion of black string perturbations in AdS space.)
The localization of black holes is also hypothesized by gravitational collapse on the brane, where ordinary matter fields are confined.

An exact solution representing a localized black hole was found in the 4D RS braneworld by Emparan, Horowitz and Myers (EHM)~\cite{Emparan:2000wa}.
However, no exact solution representing a physically acceptable localized black hole has been found in the original 5D braneworld model, although many authors have tried to construct one.
There is an argument that has been given to explain why the search for a static localized black hole solution has not yet succeeded.
It was first conjectured in Ref.~\cite{Tanaka:2002rb} on the basis of an extensive use of the AdS/CFT correspondence that localized black holes may
\textit{classically} evaporate in the 5D RS model.
If this conjecture is correct, there would be no \textit{static} solution of a black hole that is asymptotically AdS and is sufficiently large compared with the bulk curvature length.
As there exist several possibilities for stable solar mass black holes, such evaporation would set an interesting constraint on the allowed value of the bulk curvature scale.
Further discussion motivated by a no-go theorem~\cite{Bruni:2001fd} and a more explicit study showing that the existence of the EHM solution does not contradict this conjecture are found in Ref.~\cite{Emparan:2002px}. (Also see
Ref.~\cite{Duff:2000mt} for AdS/CFT in the RS braneworld.)

In contrast with the situation described above, however, recovery of 4D gravity on the 3-brane has been successfully realized in the RS models.
It was shown that 4D Einstein gravity is approximately recovered in linear
perturbations~\cite
{Shiromizu:2000wj,Garriga:2000yh,Tanaka:2000er,Mukohyama:2001ks,Tanaka:2000zv} and also in second-order perturbations~\cite
{Giannakis:2001zx,Kudoh:2001wb,Kudoh:2001kz,Kudoh:2002mn}.
Moreover, even in the case of highly relativistic stars, 4D Einstein gravity was numerically shown to be a good approximation~\cite{Wiseman:2001xt}. Hence, there are also results that suggest the existence of black hole solutions localized on the 3-brane.
If this is the case, then black holes should be produced as a result of gravitational collapse on the brane.
Because Birkoff's theorem does not hold on the brane~\cite{Bruni:2001fd}, exterior fields are not static, even during spherical collapse on the brane.
However, it is believed that the exterior fields settle down to a static state at late times within the standard picture of 4D general relativity.  This is the original interpretation of the no-go theorem~\cite{Bruni:2001fd}.

In a previous paper \cite{Kudoh:2003xz}, which is referred to as KTN in the present paper, we developed a numerical method to construct a localized black hole without assuming any artificial conditions, and we presented numerical examples of small localized black holes whose horizon radii are smaller than the AdS curvature radius.
In this paper, we study the thermodynamic properties of localized black holes using these numerically obtained solutions. We find that the numerically obtained solutions exhibit a smooth transition from the 5D Schwarzschild black hole and that the thermodynamic relations deviate significantly from those expected on the basis of naive consideration of the 5D Schwarzschild black hole.

This paper is organized as follows.
In the next section, we briefly explain the formulation for constructing the
localized black holes numerically.
In Section \ref{sec:Thermodynamical behavior}, we discuss the thermodynamic
relations obtained by analyzing the numerical solutions, and we compare them
with those for the EHM solution.
Section \ref{sec:Discussion} is devoted to summary.
The relation between the numerical solutions and the conjecture based on the
AdS/CFT correspondence is also discussed there.

\section{Numerical solutions of small localized black holes }

To obtain black hole solutions in the RS infinite braneworld model, it is
necessary to formulate the problem as a boundary value problem without assuming
any artificial boundary conditions.
In KTN, we used and developed the method that was first applied to obtain a
solution representing the gravity of a relativistic star on the
brane~\cite{Wiseman:2001xt} (see also Ref. ~\cite{Wiseman:2002zc}).
In this section, we briefly review the method and the notation presented in
KTN.

Because we wish to consider a static black hole solution that is localized on
the brane, we consider the static, $D$-dimensional axial symmetric form
\begin{eqnarray}
ds^2 = \frac{\ell^2}{z^2}
\left(
  - T^{2} dt^2 +e^{2R } (dr^2+dz^2)+r^2 e^{2C} d\Omega_{D-3} ^2
\right) \,,
\label{eq:assume}
\end{eqnarray}
where the line element $d\Omega^2_n$ is that of a unit $n$-sphere.
The cosmological constant in the bulk is related to the bulk curvature length
$\ell$ by $\Lambda= -(D-1)(D-2)/2\ell^2$, and the tension of the brane is
$\sigma=2(D-2)/8\pi G_D \ell$, where $G_D$ is the $D$-dimensional Newton constant.
If we set $T=1$ and $R=C =0$, this metric becomes the AdS metric in the
Poincar\'e coordinates.
Because this metric form has the residual gauge degrees of freedom of conformal transformations in the two-dimensional space $\{r,z\}$, we can use these conformal degrees of freedom to transform the location of the event horizon to
\begin{eqnarray}
 \rho|_{\mathrm{horizon}} = {\mathrm{const}} = \rho_h \,,
\end{eqnarray}
where we have introduced the polar coordinates
\begin{eqnarray}
&& r=\rho \sin \sqrt{\xi} \,,
\cr
&& z=\ell+\rho\cos  \sqrt{\xi}.
\end{eqnarray}

The Einstein equations yield five non-trivial equations. Three of them yield elliptic equations for the respective metric components $X_i = \{T, R, C\}$,
\begin{eqnarray}
 \triangle X_i  = S_{X_i} \,,
\end{eqnarray}
where $\triangle= \partial_r^2 +\partial_z^2 $ is the Laplace operator, and the sources $S_{X_i}$ depend non-linearly on $X_i$ and its derivatives. The remaining two equations give `constraint' equations,    $\Theta_1=0$ and
$\Theta_2=0$. All the boundary conditions that we need in order to solve the equations are derived consistently from physical requirements.
Because all the equations and all the boundary conditions can be rewritten in terms of the non-dimensional coordinates $\{\hat{t}, \hat\rho, \chi\}=\{t/\ell,
\rho/\ell, \chi\}$, we find that the system of equations is characterized by a single parameter,
\begin{eqnarray}
 L \equiv \frac{\ell}{\rho_h} \,.
\end{eqnarray}
We use this parameter to specify the black hole solutions.
In the numerical calculation, we have set $\rho_h=1$ and varied $\ell$ to
specify the value of this parameter.
The numerically obtained solutions are characterized by $L>1$, and then are small localized black holes.
Note that variation of $\ell$, keeping $L$ fixed, corresponds to a rescaling of the length scale, because the assumed metric can be rewritten as
\begin{eqnarray}
    ds^2 = \ell^2 ds_L^2\,,
\end{eqnarray}
where $ds_L^2$ is the dimensionless part of the line element given by
$\{\hat{t}, \hat\rho, \chi\}$.

\section{Thermodynamic  behavior}
\label{sec:Thermodynamical behavior}

Now we consider the 5D RS model.
In the limit  $L=\ell/\rho_h \to \infty$ for fixed $\rho_h$, the cosmological constant and the tension of the brane vanish, and hence the 5D Schwarzschild black hole is the solution in this limit:
\begin{eqnarray}
T_S = \frac{\rho^2 - \rho_h^2}{ \rho^2+\rho_h^2}\,,
\quad
R_S = C_S = \log \left(1+ \frac{\rho_h^2}{\rho^2} \right).
\label{5D Schwarzschild}
\end{eqnarray}
Thus, we see that the 5D Schwarzschild black hole is an approximate solution of very small localized black holes, and for this reason it is interesting to observe the deviations of the numerically obtained solutions from the 5D Schwarzschild black hole.
To quantify these deviations, we focus on thermodynamic quantities and elucidate thermodynamic properties of small localized black holes.

The surface gravity on the horizon, or the temperature, is determined by
\begin{eqnarray}
 \kappa = e^{-R} T_{,\rho}  \,,
\quad (  \rho=\rho_h )
\label{surface grav}
\end{eqnarray}
and the horizon volume, or entropy, of the black hole is given by
\begin{eqnarray}
 { A}_5
 =
    2 \rho_h^3\int d\Omega_2 \int^{({\pi}/{2})^2}_0
    d\xi  \frac{ \sin^2 \sqrt{\xi}}{2\sqrt{\xi}}
    \left( \frac{\ell}{z}\right)^3 e^{R+2C}  .
\label{BH area}
\end{eqnarray}
Here, the factor 2 is due to $Z_2$ symmetry.
The proper area of the intersection between the horizon and the brane is also an interesting quantity. It is given by
\begin{eqnarray}
 A_4 = 4\pi \rho_h^2  e^{2C} \,. \quad (\xi= (\pi/2)^2 )
 \label{4D area}
\end{eqnarray}
For comparison, we list the corresponding thermodynamic quantities for the 5D Schwarzschild-AdS (SA) black hole and the black string (BS):
\begin{eqnarray}
 \kappa_{\mathrm{SA}}
&=& \frac{1}{\varpi} + \frac{2\varpi}{\ell^2} \,,
\cr
{A}_{5 \mathrm{SA}} &=& 2\pi^2 \varpi^3
\,,
\cr
{A}_{4 \mathrm{SA}} &=&  4\pi \varpi^2
\,,
\label{KA5A4 for SA}
\end{eqnarray}
and
\begin{eqnarray}
 \kappa_{\mathrm{BS}} &=& \frac{1}{2\varpi}
\,,  \quad
\cr
{A}_{\mathrm{5BS}} &=& 4\pi \varpi^2 \ell
\,,  \quad
\cr
{A}_{4\mathrm{BS}} &=&  4\pi \varpi^2 .
\end{eqnarray}
Here, $\varpi$ is the circumferential radius of the horizon.
To match the conditions used in Eq. (\ref{KA5A4 for SA}) with those of Eq.
(\ref{5D Schwarzschild}) in the coordinates (\ref{eq:assume}), we must take
$\varpi = 2\rho_h$.
Note that a curvature singularity of black string appears at the AdS horizon, as well as at the ordinary black string singularity \cite{Chamblin:2000by}.

We have evaluated the thermodynamic quantities for numerical solutions. A partial data set is given in KTN.
As mentioned above, variation of $\ell$ keeping $L$ fixed corresponds to a rescaling of the length scale, and thus the length scales of the solutions differ. Hence, for comparison, we use non-dimensional combinations of the thermodynamic quantities.
Note that $\rho_h$ in our numerical calculation does not have a clear geometrical meaning, and thus it is not a relevant quantity to make a non-dimensional combination.

\begin{figure}[t]\centerline{
 \includegraphics[width=8cm,clip]{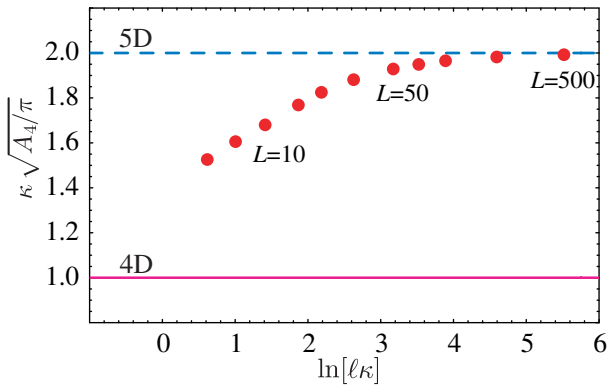}}
\caption{\label{fig:Therm A_4}
Thermodynamic relation between the surface gravity $\kappa$ and the 4D area
$A_4$.
The filled circles represent the numerical solutions for $L=
5,7,10,15,20,30,50,70,100,200$ and $500$.
For reference, we have also plotted the same thermodynamic relations for the 4D and 5D Schwarzschild black holes (the solid and dashed lines, respectively).
}
\end{figure}

\begin{figure}[t]
\centerline{ \includegraphics[width=8cm,clip]{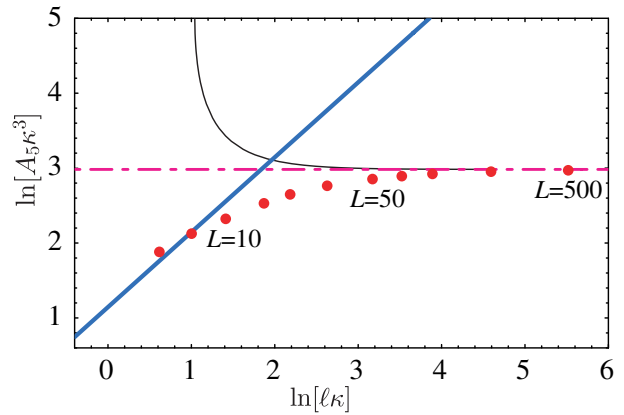}}
\caption{
\label{fig:Therm A_5}
Thermodynamic relation for the surface gravity $\kappa$ and the 5D area $A_5$ with an appropriate non-dimensional combination ($L=5$ -- $500$).
The filled circles represent the results for the numerical solutions.
The thermodynamic relations for the 5D Schwarzschild black hole (dashed line), the 5D Schwarzschild-AdS black hole (thin solid line) and the black string
(thick solid curve) are also plotted.
}
\end{figure}

In Figs.~\ref{fig:Therm A_4} and \ref{fig:Therm A_5},  we display  relations
between thermodynamic quantities.  Figure \ref{fig:Therm A_4} depicts the relation between the 4D area ${A}_4$ and the surface gravity $\kappa$, taking an appropriate non-dimensional combination.
In the figure, the thermodynamic relations for the 4D and 5D Schwarzschild black hole are also plotted for comparison.
We observe that for large $L$, i.e., for very small black holes, the numerical solutions behave like the 5D Schwarzschild black hole solution, and then, as
$L$ becomes small, they deviate from that solution and tend to behave like the
4D Schwarzschild black hole solution.
In Fig.~\ref{fig:Therm A_5},  we see the same tendency in the relation between the 5D area and the surface gravity.
When the horizon is sufficiently small ($\kappa \ell \gg 1$), the result for the numerical solutions approximately coincides with the line corresponding to the 5D Schwarzschild(-AdS) black hole.
However, as the horizon radius increases, the numerical solutions begin to deviate from this line and move to the direction along the black string.

\begin{figure}
\centerline{\includegraphics[width=8cm,clip]{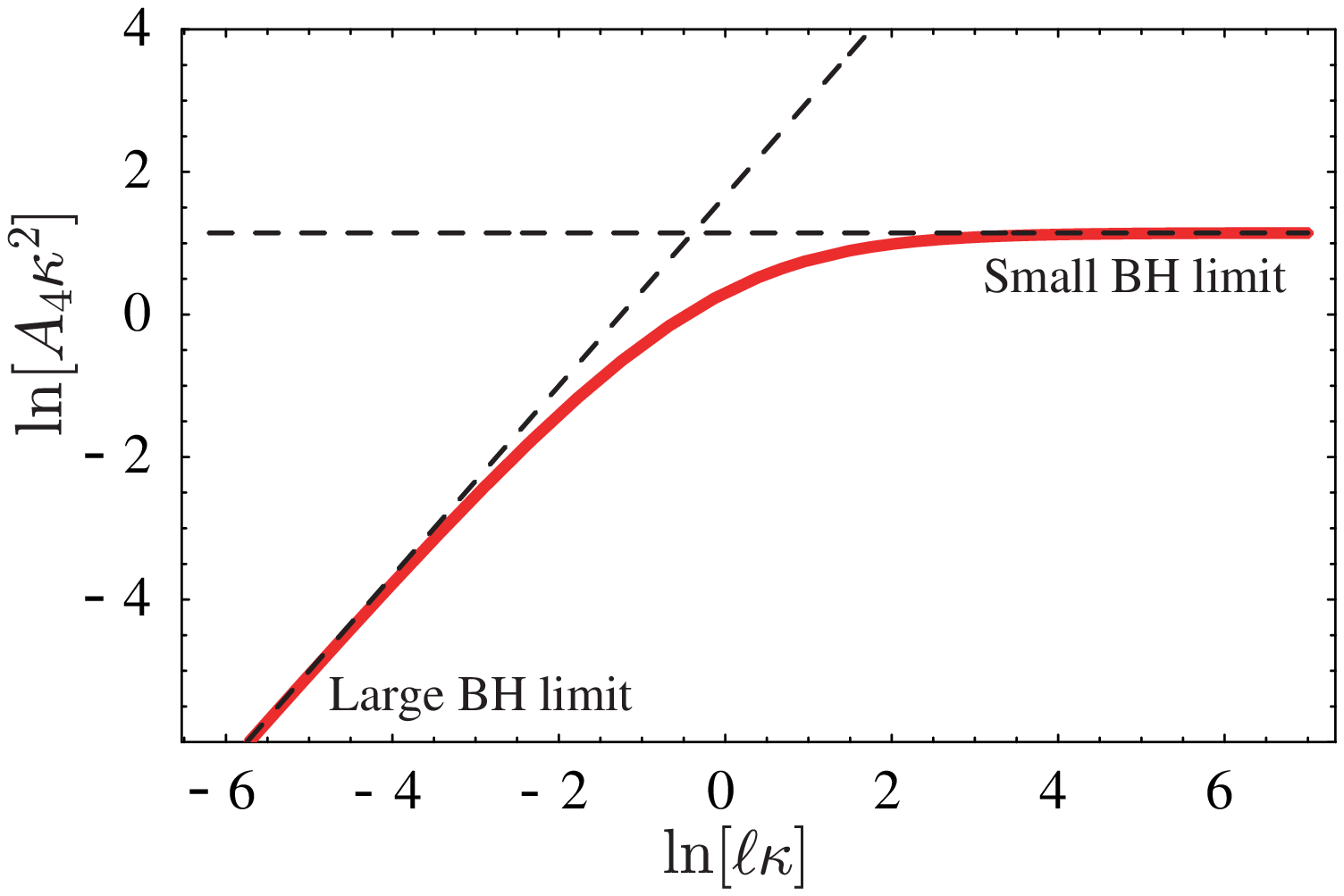}}
\caption{
\label{fig:4D Exact}
Thermodynamic relation for the EHM solution in the 4D braneworld model
\cite{Emparan:2000wa} (see Fig. \ref{fig:Therm A_5}).
The two dashed lines represent the small and large limits of the black hole,
which are given by $A_4\kappa^2=\pi$ and $A_4\kappa^2= (2^{7/3}\pi /3)
(\kappa\ell)^{4/3}$, respectively.
}
 \vspace{1cm}
\centerline{\includegraphics[width=8cm,clip]{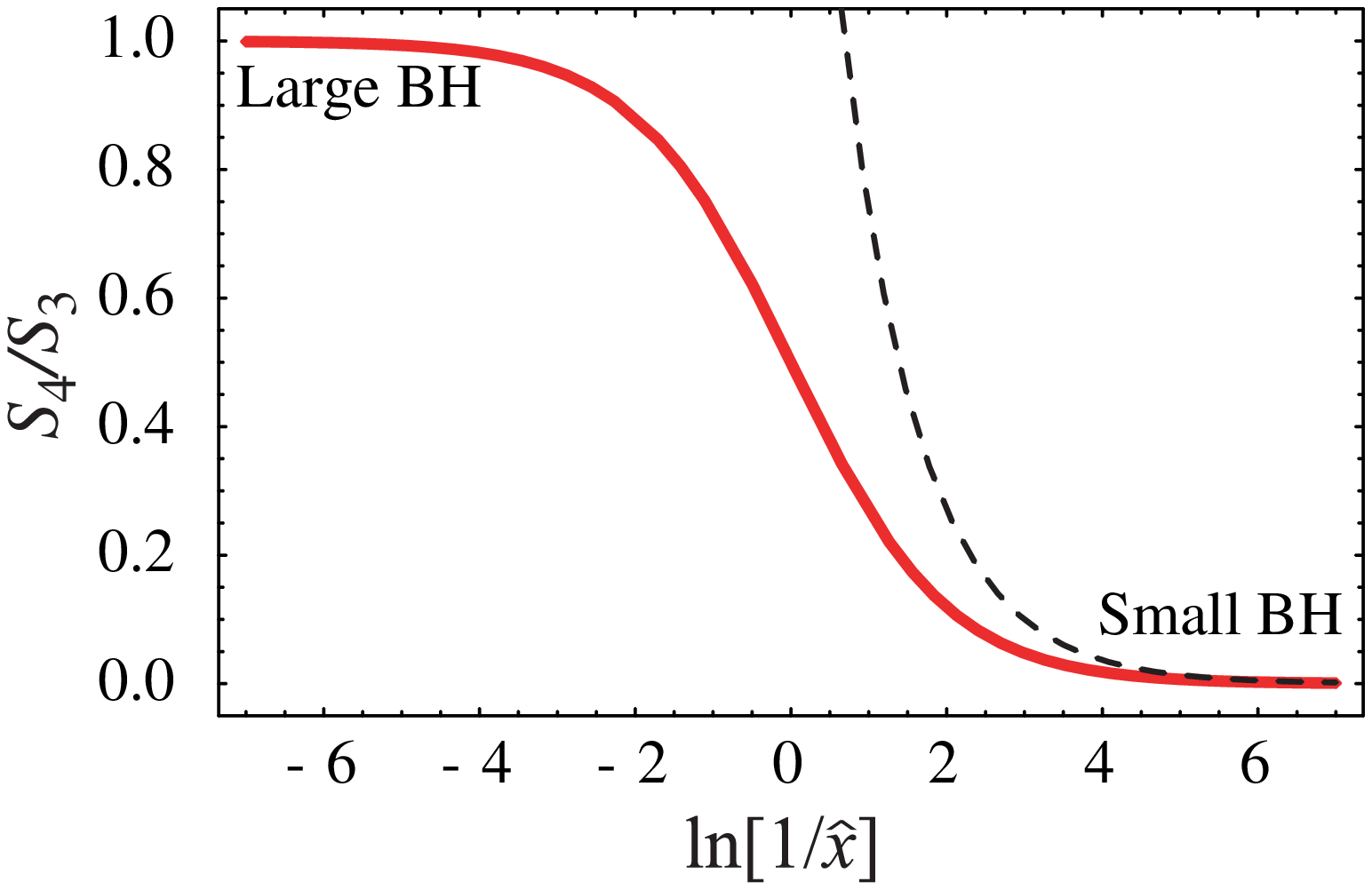}}
\caption{
\label{fig:EntropyRatioEHM}
Ratio of the 4D entropy $S_4$ to the 3D entropy $S_3$ for the EHM solution
(see Fig. \ref{fig:EntropyRatio}).
 $\hat{x}$ specifies the mass parameter of the solution\cite{Emparan:2000wa}.
For comparison, we also plot the ratio for the 4D Schwarzschild black hole
(dashed curve).
}
\end{figure}

The EHM solution~\cite{Emparan:2000wa} is the exact black hole solution in the 4D braneworld.
However, because of the low dimensionality, the black hole is not asymptotically AdS, but has a deficit angle, and it is uncertain whether the properties of the EHM solution reflect general characteristics of a black hole in the original 5D braneworld. Therefore it is interesting to compare the thermodynamic relations for the 5D localized black hole with the relations for the EHM solution.
We find similar behavior of these thermodynamic relations for $\ln \kappa \ell
\gtrsim 0$ in Fig.~\ref{fig:4D Exact}, and this shows that the 4D exact solution gives a good description of the qualitative behavior of the thermodynamic relations for the 5D solution, at least for small horizon radii.
In addition, this seems to give insight into a method for extrapolating the thermodynamic relations to larger black hole solutions.
However, for large black holes, there is also a big difference between the
5D model and the 4D model, as discussed in Refs.~\cite{Tanaka:2002rb} and
\cite{Emparan:2002px}.
On one hand, the metric induced on the brane for a large black hole in the 5D model is conjectured to mimic the 4D Schwarzschild metric with some small
corrections. On the other hand, there are no 3D black hole solutions without a cosmological constant that correspond to the 4D Schwarzschild black hole in the 5D model.
Thus, we need to be careful with speculation based on analogy to the EHM solution.

\begin{figure}
\centerline{\includegraphics[width=8cm,clip]{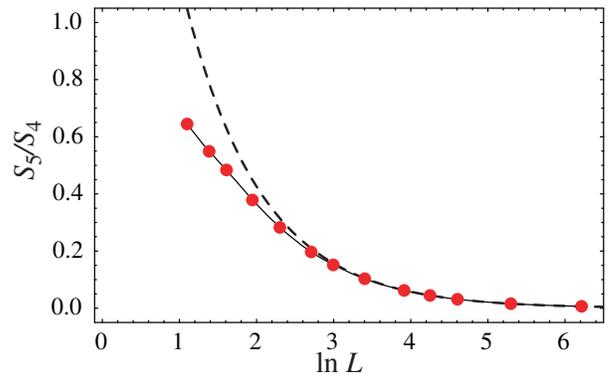}}
\caption{
\label{fig:EntropyRatio}
Ratio of the 5D entropy $S_5$ to the 4D entropy $S_4$.
The dashed curve represents the ratio for the 5D Schwarzschild black hole.
 }
\end{figure}

Let us consider the shape of a small localized black hole in the bulk.  In KTN,
we derived the ratio of the mean radius in four dimensions,  $\sqrt{A_4}$, (on
the brane) to that in five dimensions, $A_5^{1/3}$,  indicating that such a
black hole tends to flatten as its horizon radius increases.
This flattened shape is understood to be due to the so-called warp factor,
$\ell z^{-1}$, of AdS geometry (\ref{eq:assume}).
In Fig.~\ref{fig:EntropyRatio}, we display the ratio of the 5D entropy
$S_5=A_5/4G_5$ to the 4D entropy $S_4=A_4/4G_4$. Here, we use the fact that
Newton constant $G_5$ in five dimensions is related to that in four dimensions
by $G_5=\ell G_4$. This ratio also gives information on the shape in the bulk.
If the black hole horizons have an almost flattened shape, extending roughly a
distance $\ell$ off of the brane, the ratio becomes $\approx 1$. In fact, the
corresponding ratio of the EHM solution exhibits such behavior
(Fig.~\ref{fig:EntropyRatioEHM}).
Figure~\ref{fig:EntropyRatio} shows a similar tendency. However, we could not
confirm if the ratio $S_5/S_4$ approaches some constant value, or if it
increases, without approaching a value.

\begin{figure}
\centerline{\includegraphics[width=8cm,clip]{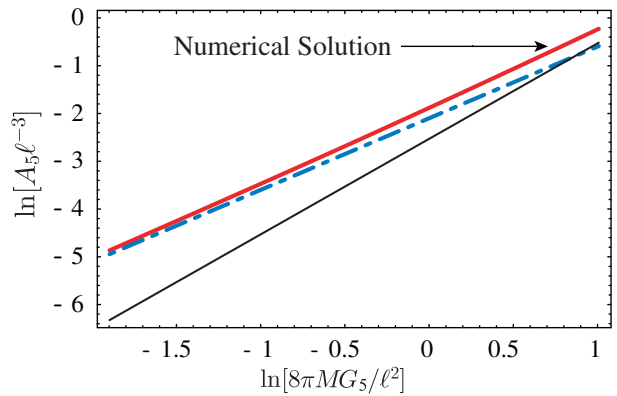}}
\caption{
\label{fig:MassEntropy}
The relation between the thermodynamic mass and the entropy ($L=5
\sim 500$).
The dashed line and the thin line represent the 5D Schwarzschild black hole and the black string, respectively.
 }
\end{figure}

The mass of a black hole is one of its important quantities, and it would be
interesting to determine the mass of localized black holes.
However, it is difficult to extract the mass asymptotically, for example on the
brane, from numerical solutions. To obtain an estimate of the mass, we simply
assume the first law of thermodynamics, represented by  $dM= \kappa dA/8\pi
G_5$, and we calculate the thermodynamic mass $M$ by integrating this quantity.
In order to carry out this integration, we interpolated the thermodynamic
quantities obtained for discrete values of $L$ up to $L=5$. The smallest
numerically obtained black hole ($L=500$) was smoothly extrapolated to the 5D
Schwarzschild black hole. Figure \ref{fig:MassEntropy} displays the
mass-entropy relation. For comparison, the mass-entropy relations for the black
string and the 5D Schwarzschild black hole are also plotted.
 We see that the localized black hole ($L \gtrsim 5$) has greater entropy, at
least in the plotted region, than both the black string and the 5D
Schwarzschild black hole of the same mass.
 Extrapolating the curves, however, it seems that the entropy of the black
string is likely to become larger than that of the localized black hole as the
mass increases. If this is the case, the localized black hole solution would
seem to be unstable from the point of view of the entropy.
Of course, we know that the black string in the RS braneworld is dynamically
unstable, as long as we add a negative tension brane at some finite but
sufficiently distant position \cite{Gregory:1993vy,Gibbons:2002pq}, and
therefore the larger entropy does not imply the dynamical stability of the
black string. This observation might be very suggestive, but we cannot give any
conclusive statement about the dynamical stability of the localized black hole,
because stability based on entropy does not lead directly dynamical stability.

\section{Discussion}
\label{sec:Discussion}

Physically acceptable black hole solutions that represent small black holes
localized on the brane in the RS infinite braneworld model were constructed in
an approximate manner using a numerical method in KTN~\cite{Kudoh:2003xz}.
The horizon radius $\rho_h$ of the numerically obtained black holes is small
compared to the bulk curvature scale $\ell$.
In this paper, we have studied the thermodynamic properties of small localized
black holes and found that the thermodynamic relations of the numerically
obtained solutions deviate significantly from those expected on the basis of
naive consideration of the 5D Schwarzschild(-AdS) black hole as its horizon
radius increases (Fig. \ref{fig:Therm A_5}).
Comparison of the thermodynamic quantities also suggests that the numerical
solutions approach solutions whose induced metric on the brane behaves like
that of the 4D Schwarzschild black hole  (Fig.~\ref{fig:Therm A_4}), although
we could not obtain clear evidence for this from the direct comparison of the
induced metric.
Moreover, small localized black holes have greater entropy than a black string
of the same thermodynamic mass (Fig. \ref{fig:MassEntropy}), and a localized
black hole undergoes a shape  transition in the bulk
(Fig.~\ref{fig:EntropyRatio}, and see also KTN). A black hole tends to flatten
as its horizon radius increases.

Although the EHM solution is the solution in lower dimensions, it is
interesting to compare our solution with it.
 We have compared the thermodynamic relations of the EHM solution with those of
the 5D numerical solutions and observed similarities between them (Figs.
\ref{fig:4D Exact} and \ref{fig:EntropyRatioEHM}).
This observation might be suggestive, but because our calculations are
restricted to only small black hole solutions ($\rho_h/\ell < 1$), we cannot
state a definite conclusion regarding large black hole solutions ($\rho_h/\ell
\gtrsim 1$).
Thus, it is an important problem to determine the nature of large black hole
solutions. In particular, induced gravity and its deviation from the 4D
Schwarzschild black hole at large distances are interesting.

Here we discuss the relation between our results and the classical black hole
evaporation conjecture~\cite{Tanaka:2002rb,Emparan:2002px}.
Based on the AdS/CFT correspondence, it was argued that there might be no
static black hole solution in the RS single brane model.
It might be thought that the discovery of small black hole solutions
contradicts this conjecture.
However, there are two possible consistent scenarios.
One is that only small black hole solutions exist in the strict sense.
Because the classical evaporation conjecture applies only for sufficiently
large black holes, for which the quantum correction in the AdS picture is negligible, the existence of small black hole solutions does not directly contradict the conjecture.
However, the relations between thermodynamic quantities suggest that there is no critical point at which the sequence of solutions suddenly ceases to exist.
Therefore, we think that this first possibility is less likely.
The other possibility is that even small black hole solutions exist only in an
approximate sense.
Even if there is no static black hole solution in the strict sense,
irrespective of its size, the internal inconsistency contained in the setup of
the problem might be made irrelevant by numerical errors.
If this is the case, the failure to find larger black hole solutions might be a
signal of increased effect of the inconsistency.
Of course, there also remains the possibility that the classical evaporation
conjecture is incorrect.
Unfortunately, however, we cannot state anything definite about these points at
the moment, because the lack of convergence we encountered in the relaxation
scheme could be a purely technical problem with the numerical scheme
\cite{Kudoh:2003xz}.

We have to this point discussed the 5D numerical solutions obtained in KTN. 
The same formulation and numerical method can be applied to obtain localized black holes in higher dimensions ($D\ge 5$).
Numerical calculation in higher dimensions has an advantage
\cite{Wiseman:2002zc}: The metric in 6 dimensions, for example, dies away faster than that in 5 dimensions, and therefore an asymptotic boundary condition can be imposed at a location closer to the center of the simulation volume. 
This would reduce significantly the total lattice size, and hence make the numerical calculation more tractable. Actually, using the same numerical scheme, we have performed the calculation in 6 dimensions. The result of the numerical calculations shows that 6D localized black holes have thermodynamic properties similar to those of 5D black holes. Furthermore, we cannot again maintain the convergence of the calculation for $L \lesssim 1$, and only small localized black holes are found.
There are many open questions regarding the features of localized black holes.
To obtain a thorough understanding of localized black holes, it is crucially important to find large black holes, if they exist.
In order to realize this, we need further development of investigational methods.

\begin{acknowledgments}
The author would like to thank Takahiro Tanaka and Takashi Nakamura for their valuable comments and suggestions.
To complete this work, discussions during and after the YITP workshops
YITP-W-01-15 and YITP-W-02-19 were useful.
The numerical computations reported in this work were carried out at the Yukawa Institute Computer Facility.
This work is supported by the JSPS.
\end{acknowledgments}


\end{document}